\documentstyle[12pt]{article}
\begin{document}
\title{\bf Do Hard Spheres have Natural Boundaries?}

\author{Barry M. McCoy{\footnote{e-mail:mccoy@insti.physics.sunysb.edu}}\\
Institute for Theoretical Physics\\
State University of New York\\
Stony Brook, N.Y. 11794}

\maketitle

\begin{abstract}

I use recent advances in the study of the susceptibility of 
the Ising model to propose a new mechanism for the freezing transition
which is observed in three dimensional hard spheres.
\end{abstract}

\section{Introduction}

Recently it has been shown \cite{n1} that the susceptibility of the two 
dimensional Ising model at $H=0$ has a natural boundary in the complex 
temperature plane. Furthermore the first 323 terms in  the high 
and low temperature series expansions have 
been generated \cite{ongp1},\cite{ongp2}. This discovery of a natural boundary 
gives new insights into phase transitions. In  this note I explore possible 
consequences for the phenomenon of hard sphere freezing.

The freezing transition in the classical hard sphere system has 
been observed in both molecular dynamics  and 
Monte Carlo computations since the 50's \cite{wood},\cite{alder} and 
60's \cite{hoover}, \cite {alder2} and by now it is well established \cite{lowen} that
there is a first order transition with the coexistence of a fluid phase with a 
dimensionless volume fraction of $\eta_f=0.494$ and a solid phase with a volume fraction of 
$\eta_s=.545$ where $\eta$ is defined as 
$\eta=\rho v_{hs}=\rho{4\over 3}\pi r^3_{hs}$ with $r_{hs}$ 
the radius of the hard spheres and $\rho$ is the number density $N/V.$ 
A non equilibrium glass transition occurs at $\eta_g=0.58.$ 
By comparison the volume fraction of the closest 
packed configuration (the FCC lattice) is 
$\eta_{cp}={\pi\over 3\sqrt 2 }=0.7405$ which was conjectured 
by Kepler in 1611 and recently proven by Hales \cite{hales}. There is 
also a volume fraction $\eta_{rcp}=0.638$ at which random 
close packing occurs.  

Because the volume fractions $\eta_f$ and $\eta_s$ lie well 
below the densities $\eta_g$ and $\eta_{rcp}$ where possible 
non ergodic behavior
might occur the freezing transition is expected to be described 
by the partition function of the canonical ensemble. Yet , somewhat 
surprisingly, after more than three decades of effort the existence 
of this transition is still considered controversial.
This is very well stated in a recent paper \cite{kegel} ``
Theory can explain the transition only ``afterwards'', that is, 
properties of the coexisting phases should be 
inserted into the theories in advance, and the 
coexisting densities follow by minimizing the free energy functional''.   

The theory which is usually used to describe the transition (see 
for example ref.\cite{lowen} and references contained therein) has remained 
virtually unchanged for over 30 years \cite{hoover},\cite{alder2}.
Typically the high density phase is described by a cell (or free volume) 
approximation 
\cite{munster} which has the virtue that it is the exact limiting form 
of the free energy as $\eta\rightarrow \eta_{cp}.$ The 
fluid phase is described by solving a liquid integral equation \cite{hansen} 
based on the Percus--Yevik approximation    or the scaled 
particle approximation \cite{reiss}
to the fluid free energy. 
A more modern approach uses density functional theory \cite{evans}.
 These high and low density expressions are then extrapolated out of 
their regions of validity and in this extrapolated region the densities of 
coexisting phases are found by a Maxwell construction. This procedure is 
guaranteed to give a first order transition but cannot actually be called 
predictive. 

On the other hand this description of a first order transition
 has long ago been called 
into question by the droplet models of condensation of Fisher \cite{fisher} 
 and Langer \cite{langer}
which predict  an essential singularity at 
the coexistence density. 

These two points of view of first order transitions are not 
in harmonious agreement and the state of 
knowledge is perhaps well described by ref.\cite{kegel} 
``$\cdots$ from a fundamental viewpoint they [the existing theories of 
freezing] 
are still unsatisfactory. The splitting of a hard sphere system into a fluid 
and a solid branch is, evidently, hidden in the partition function of the
system. A thorough theory of hard sphere freezing should therefore 
identify this property that leads to the observed symmetry breaking''

The purpose of this note is to use the intuition obtained 
from the recent Ising model computations \cite{n1}-\cite{ongp2} 
to reexamine the hard sphere freezing transition and to propose that the 
phase transition is due to
the formation of a natural boundary in the free energy 
in the complex density plane.
  
\section{Natural Boundaries}

Yang and Lee \cite{ly} pioneered the study of phase transitions by means of 
looking at the zeroes of the partition function on a finite size lattice.
These zeros will in general not be on the positive real  axis 
(fugacity, temperature or density) but may approach the axis in the limit 
as $V\rightarrow \infty.$ We may classify  several of the possible 
limiting behavior of zeroes as follows:

1) The zeroes can lie on a curve in the complex plane and pinch the real axis 
at a point (say $T_c$ for example). In the limit $V\rightarrow \infty$  
the curve of 
zeroes in the complex plane becomes a branch cut and the point of pinching 
becomes a branch point. This happens in the complex temperature 
plane for all integrable models and forms the basis of the theory of 
second order phase transitions. In this picture the thermodynamic functions 
are singular only at $T_c,$ the point of phase transition.

2) The zeroes can lie in some area of the complex plane,   
pinch the real axis in some line segment and cut 
the plane into two 
disconnected pieces each of which may be analytically continued
into the region occupied by the zeroes in the finite size system.
In this picture the thermodynamic functions such as the free energy 
and the pressure are analytic at the coexistence density $\eta_f$
and the virial expansion will converge for densities greater than $\eta_f.$
 This  picture forms the basis of the theory of first order 
phase transitions and is exactly the opposite of what happens 
in a second order transition.

3) The zeroes may pinch the real axis at either a point or a line segment 
and form a natural boundary in the 
complex plane. This is the phenomena discovered by Orrick, Nickel, 
Guttmann and Perk \cite{n1}-\cite{ongp2} in the susceptibility 
of the Ising model.

The first mechanism of phase transition is explicitly seen in every
integrable model whose partition function has been computed. However, the 
computations of ref. \cite{n1}-\cite{ongp2} strongly lead to the conjecture 
that for non integrable models with a second order phase transition there 
will always be a natural boundary accompanying the singularity at $T_c$ which
gives the usual critical exponents. In this sense it may be expected that 
natural boundaries are generic in second order transitions  and 
that it is only the very special symmetries of integrablity
which turn the curve of zeroes into a branch cut in the case of 
integrable models.

For first order transitions such as hard spheres at the freezing density 
the situation is much less clear. First of all it may be debated whether a 
freezing transition in which the system undergoes a change in symmetry 
is the same as the vapor/fluid condensation where there is no symmetry 
change. Even for the liquid/vapor transition   
the only models which have ever been solved are mean 
field models with long range attractive forces such as the Kac
potential model  of condensation \cite{hl}. The pressure 
of these models is in fact analytic at the coexistence curve.
However, for the Ising model at sufficiently 
low temperatures  the free energy is not analytic at zero magnetic field
\cite{isakov} which in the language of the lattice gas means 
that there in a singularity at the coexistence boundary.  
This singularity is of the infinitely differentiable type discussed by Fisher 
\cite{fisher} and  Langer \cite{langer} but the
additional prediction that the singularity is an isolated essential 
singularity has not been addressed.

Furthermore no model in a finite number of dimensions with only repulsive 
forces has ever been solved exactly  which exhibits a freezing transition.    

There is thus no compelling evidence to support the assumption that the 
scenario of case 2 must be correct for hard spheres and 
thus it is possible to raise the suggestion that case three  holds 
for first as well as second order transitions. The difference 
being that for first order transitions the behavior at the 
coexistence density is presumably infinitely differentiable (but of 
course not analytic) instead of having an algebraic singularity.

This conjecture can be proven wrong if it can be demonstrated 
that there is a lower bound on the radius of convergence of the 
virial expansion which is greater than the fluid coexistence density of 
$\eta_f=0.494.$ However, the existing bound  Lebowitz and Penrose \cite{joel}
$\eta_{pl}=.01809,$ which comes from Groeneveld's bound \cite{hans} on the 
fugacity expansion of the pressure in the grand canonical ensemble , 
is very far less than this density 
so the conjecture cannot be ruled out by the existing bounds.

It is of course widely appreciated that the bound $\eta_{lp}$ is much 
less than $\eta_f.$ However, because of the excellent agreement of 
the Pad{\'e} approximate extrapolation of the first eight virial coefficients 
with the numerical simulations it is often said that ``$\cdots$ 
it (the lower bound of Lebowitz and Penrose) seems to be far 
below the true radius of convergence'' \cite{janse}. Indeed the nearest 
singularity in the Pad{\`e} approximations is almost exactly 
at $\eta_{cp}$ and no sign of any kind is seen of the freezing 
transition in the eight term virial expansion data of ref. \cite{janse}.

It is exactly at this point that the intuition obtained from the natural 
boundary in the Ising model is extremely useful. In ref. \cite{n1} the 
existence in the susceptibility of a dense set of singularities
lying on the curve of zeroes of the partition function is demonstrated. But 
in refs. \cite{ongp1}-\cite{ongp2} where an expansion of the susceptibility is 
made to 323 terms only a small number (up to 8) of these dense set of 
singularities can be resolved out of the series expansion. 
We therefore conclude that unlike
the power law singularities of second order transitions which are very strong 
and can be located with series as short as 8 or 10 terms that the 
singularities coming from natural boundaries require series of hundreds or 
thousands of terms to see. It is thus completely plausible and possible that
the Pad{\'e} approximate to the 8 term virial expansion is incapable of 
locating a natural boundary singularity at $\eta_f$ even if it does exist.
 Indeed it may be argued that since the kissing 
number (maximum number of spheres which can touch a given sphere)
is 12 in three dimensions that the virial expansion cannot possibly  
include the effects of the geometry of hard spheres until at least the 12th 
coefficient has been computed. 
This is reminiscent of the early studies \cite{sn} on the Ising susceptibility
where, based on the first 8 terms of the series expansion, 
a simple algebraic expression was 
conjectured which in fact omits almost all of the interesting structure of
the true answer.

To illustrate the difficulty in seeing natural boundaries in a virial 
expansion consider the familiar example of the 
lacunary series 
\begin{equation}
\sum_{k=1}^{\infty}a_kz^{n_k}~~
{\rm with}~~{\rm lim}_{k\rightarrow \infty}k/n_k=0,
\label{lac}
\end{equation}  
which is well known to have a natural boundary.
Typical examples of $n_k$ are $k^2$ and $2^k.$ The rapid growth of the powers
of $z$ indicate that the natural boundary will be invisible in a Pade analysis 
of terms up to order $z^8.$
Note that this example exhibits the general  feature of 
natural boundaries is that there 
is a sense in which  the coefficients in the power series have 
oscillations which in a Pad{\'e} analysis will lead to the appearance of 
more and more singularities in the complex plane as the number of terms 
in the series increases. We also note that if the coefficients 
$a_k$ fall off sufficiently rapidly (say as $1/k!$ for $n_k=k^2$ or $2^k$) 
that the singularity 
at $z=1$ will be infinitely differentiable. This is the behavior found by 
Isakov for the Ising model \cite{isakov} and we therefore conclude that this 
result is just as compatible with a natural boundary as it is with the
essential singularity of Fisher \cite{fisher} and Langer \cite{langer}.

It has, of course, always been possible to assert that the convergence of 
the virial expansion is determined by terms which have not yet been seen 
and without some evidence to the contrary it is just as reasonable to 
ignore this possibility. What has changed with the discovery of the 
natural boundary in the Ising model susceptibility is that not only do we now 
have an example where a natural boundary does occur but the reason which 
it occurs 
seems to be connected with the fact that at $H\neq 0$ the Ising model 
is not integrable. Therefore it would seem  that natural boundaries are
features of generic models (such as hard spheres) instead of the very 
specialized models such as the Ising model at $H=0$ 
 
Finally it is perhaps useful to comment on the question of 
possible sign changes in the virial coefficients.
The mechanism of natural boundary formation can happen if all the 
virial coefficients are positive as the example of the lacunary series 
(\ref{lac}) shows. However it has been suggested as long ago as 1957 by 
Temperley \cite{temp} that `` 
the possibility of violent oscillations [in the signs of 
the virial coefficients]
cannot altogether be ruled out.''
Furthermore Ree and Hoover \cite{rh} showed in dimension 8 and 9 
that  that the 
fourth virial coefficient is indeed negative but they 
speculated that in three dimensions the signs would not 
change before the ninth or tenth coefficient.
Indeed the first eight 
coefficients for hard spheres and discs \cite{janse} are now known to
be   positive
as is the Pade estimate for the ninth coefficient. 
Nevertheless Temperley's suggestion of violent oscillations in sign is an 
appealing mechanism because random oscillations in sign almost always 
lead to natural boundaries.

\section{Conclusions}

The possibility of a natural boundary in the pressure as a function of density
as an explanation of the freezing transition of hard spheres
is unorthodox but 
it does in fact have a certain intrinsic charm. Not only does it put 
first and second order transitions in some sense on an equal footing    
but it gives some hope that there is some predictability to 
first order transitions. If the mechanism of first order transitions is in 
fact that there are 2  (or more) branches of the free energy which 
do not know about each other then it is extremely difficult to 
ever say if you have in fact complexly computed the phase diagram. 
There may always be some other phase with a lower free energy 
which you just never thought of. In this situation one will 
always be in the position described by ref. \cite{kegel} that ``theory 
can only explain the transition afterwards''. In the end this 
reduces statistical mechanics to a descriptive rather than predictive 
science. Indeed this may account for the fact that many physicists prefer 
to study second order transitions and are happy to leave 
first order transitions to chemists and biologists. If, on the other hand, 
the natural boundary mechanism proves to be correct then the existence of 
the phase boundary can at least in principle be determined by studying the 
properties of the fluid phase alone. In many ways this is a much more 
attractive  alternative. At the very least sufficient analysis 
should be done on the hard sphere system to demonstrate 
that it does not happen.

\vspace {.2in}

{\Large \bf Acknowledgments}

\bigskip
The author wishes to thank A. Guttmann, W. Orrick and 
J.H.H. Perk for discussions about their work on the Ising susceptibility. He 
also is indebted to R. Kamien , T. Lubensky and A. Sokal 
for extremely  illuminating discussions about hard spheres.
This work is supported in part by NSF grant DMR0073058.

\end{document}